\documentclass[%
 preprint,
 superscriptaddress,
 amsmath,
 amssymb,
 showkeys,
 aps,
]{revtex4-2}

\usepackage{graphicx}
\usepackage{dcolumn}
\usepackage{bm}
\usepackage{hyperref}
\usepackage[mathlines]{lineno}
\usepackage{array}
\usepackage{commath}
\usepackage{booktabs}
\usepackage{lipsum}

\usepackage{hyperref}
\hypersetup{
    colorlinks=true,
    citecolor = red,
}

\usepackage{etoolbox}
\makeatletter
\patchcmd{\frontmatter@abstract@produce}
  {\vskip200\p@\@plus1fil
   \penalty-200\relax
   \vskip-200\p@\@plus-1fil}
  {}
  {}
  {}
\makeatother

\begin{document}
\title{Transportation Interventions Reshaping NYC Commute: the Probabilistic Simulation Framework Assessing the Impacts of Ridesharing and Manhattan Congestion Surcharge}

\author{Devashish Khulbe}
\thanks{Devashish Khulbe and Chaogui Kang contributed equally to this work.}
\affiliation{%
 Center for Urban Science + Progress, New York University, Brooklyn, NY 11201, USA
}%

\author{Chaogui Kang}
\email{cgkang@whu.edu.cn}
\affiliation{%
 School of Remote Sensing and Information Engineering, Wuhan University, Wuhan, Hubei 430079, China
}%
\affiliation{%
Center for Urban Science + Progress, New York University, Brooklyn, NY 11201, USA
}%

\author{Stanislav Sobolevsky}
\email{ss9872@nyu.edu}
\affiliation{%
 Center for Urban Science + Progress, New York University, Brooklyn, NY 11201, USA
}%

\date{\today}

\begin{abstract}
Understanding holistic impact of planned transportation solutions and interventions on urban systems is challenged by their complexity but critical for decision making. The cornerstone for such impact assessments is estimating the transportation mode-shift resulting from the intervention. And while transportation planning has well-established models for the mode-choice assessment such as the nested multinomial logit model, an individual choice simulation could be better suited for addressing the mode-shift allowing to consistently account for  individual preferences. In addition, no model perfectly represents the reality while the available ground truth data on the actual transportation choices needed to infer the model is often incomplete or inconsistent. The present paper addresses those challenges by offering an individual mode-choice and mode-shift simulation model and the Bayesian inference framework. It accounts for uncertainties in the data as well as the model estimate and translates them into uncertainties of the resulting mode-shift and the impacts. The framework is evaluated on the two intervention cases: introducing ride-sharing for-hire-vehicles in NYC as well the recent introduction of the Manhattan Congestion Surcharge. Being successfully evaluated on the cases above, the framework can be used for assessing mode-shift and resulting economic, social and environmental implications for any future urban transportation solutions and policies being considered by decision-makers or transportation companies. 
\end{abstract}


\maketitle

\newpage
\section*{Introduction}

The vast scale of New York City (NYC) can magnify even a slight improvement in the efficiency of the transportation solutions translating it into huge cumulative economic, environmental and societal impacts. The rapidly growing for-hire vehicles (FHV) service is one area which can realize such optimization of drastically improving the efficiency of car and taxi transportation, as intended to cut traffic, congestion and energy consumption \cite{santi2014carpool}. The surge in ride-sharing trips in the recent years have demonstrated that the FHV service is playing an increasingly important role in the city's overall transportation (over 2.5 times from mid-2017 till end of 2018 with over 25 million miles traveled monthly by the end of 2018 on the shared rides according to NYC TLC open data) \cite{palombo2019taxi}. Such potential has been further unleashed with an in-depth understanding of the basic urban quantities/parameters (such as city size and driving speed) that affect the fraction of individual trips that can be shared \cite{tachet2017shareability}. Unfortunately, modal shifts resulting from increased affordability of the FHV service can easily offset those positive impacts, contributing to a substantial proportion of the overwhelmed road traffic and energy emission. Another issue resulting from the growing number of vehicles is the increased traffic congestion in the city. Both citywide bus speeds and the average travel speed within Manhattan's central business district (the area south of 60th Street) are the slowest they have been in decades (buses average 7.58 miles per hour—it was 8 miles per hour in 1990—while the travel speed in Manhattan is now just over 7 miles per hour, down from 9 miles per hour in 1990) \cite{NYCmobilityReport}. In addition, close to 45 percent of New Yorkers get a delivery at home once per week, which not only affects how many trucks are on city streets, but how vehicles can get around. The city is putting congestion pricing as one of the measures into place that may combat these problems. Meanwhile, a discrimination of the impacts against different transport alternatives and for different population groups with distinct demographics is essential \cite{kodransky2014income}. Urban stakeholders and municipal managers need to make informed decisions while considering policies and adopting solutions based on the travel behavior simulations driven by such knowledge, ideally, in a social petri dish. 

The behavioral framework and a model for the set of complete and inter-related choices undertaken by travelers and potential travelers in the travel market is required for such assessments. Both aggregate and disaggregate approaches have been developed to estimate travel demand and to split modal choices \cite{koppelman2006choice}. Those popular and widely-used include the Gravitational models \cite{anas1983gravity}, the Probit models \cite{alemi2019probit}, the Logit models \cite{wen2001logit} and many others. The explanatory variables included in the models often involve demographic, socioeconomic character, trip characters and mode attributes \cite{wen2001logit, scheiner2007life}. In particular, many empirical evidences have suggested that, among the developed models, the Logit model often has more analytical advantages and offers more accurate results \cite{ghareib1996logit}. It has been taken as a reference model for urban travel mode choice simulations for a long time. Given a set of alternatives available, the probability of selecting a mode is determined by the Logit model (or its variants) as a function of the systematic portion of the utility of all the alternatives \cite{wang2017probability}, and the proportion of modal shift under intervention scenarios can also be determined \cite{tuite2014shift}. The most common model structures are the multinomial logit and nested logit models, which assume that the alternatives are grouped in nests (or combined modes) and the alternative (within each combined mode or between different combined modes) are independent from each other \cite{bekhor2010nmml}. For the estimation and evaluation of a practical mode choice model, traveler and trip related data including the actual mode choice of the traveler are required, which should be obtained by surveying a sample of travelers from the population of interest.

A citywide synthesis at a mega-scale like NYC requires enormous amount of detailed travel information for the modeling to understand the factors that affect travel-related choices and to predict how people travel in time and space. For decades, transportation researchers have largely used survey data of active solicitation \cite{chen2016bigdata}, which are detailed but limited by a relatively small sample size (small data). The rapid rise and prevalence of mobile technologies have enabled the collection of a massive amount of passive data (big data) very different from data of active solicitation (small data) that are familiar to most transportation researchers and requires different methods and techniques for processing and modeling \cite{gonzalez2008scaling, liu2015social, yue2014review, ukkusuri}. In recent years, data on human mobility and interactions (such as proxies from anonymized cell phone connections \cite{girardin2008digital, gonzalez2008scaling, amini2014impact, kung2014exploring, grauwin2017}, credit card transactions \cite{sobolevsky2014money, sobolevsky2016prism}, GPS readings \cite{santi2014carpool, nyhan2016, qian2019}, geo-tagged social media \cite{hawelka2014, paldino2015, belyi2017} as well as various sensor data \cite{const2016}) in the city space saw an increasing number of applications. 

A critical drawback lies in having the available data either not including any user demographic information for individual trips, or providing travel statistics with demographic information at the aggregate level only, as a response to alleviate privacy and surveillance concerns \cite{douriez2016privacy}. A synergy of disclosed (small and big) travel data from different data providers and departments is often required \cite{huang2018fusion, li2019resample, beiro2016assimilation}: \emph{to represent the resultant complete travel information (such as number of trips, travel time, and monetary cost) at a certain aggregate level, and it, subsequently, is not as accurate and detailed as the unsynergized incomplete data}. Such compromise imposes uncertainties onto both the data reliability and the modeling process \cite{manzo2015uncertainty, trajcevski2011uncertainty, rasouli2012uncertainty}, suggesting that the point estimates of modelled modal choices only represent one of the possible outputs generated by the models and, instead, anticipated modal choices are better expressed as a central estimate and an overall range of uncertainty margins articulated in terms of output values and likelihood of occurrence \cite{boyce1999uncertainty}. 

In this article, we seek building a methodological framework to explore the modal choice behaviors in NYC using a data-driven framework based on partial ground truth data and with consideration of both data and model uncertainties. By evaluating the synthesized transportation choices under scoping scenarios as well as the actual up-to-date taxi and FHV ridership, we train the mode-choice simulation model capable of simulating further mode-shift on the individual level under intervention scenarios of interest - introduction of ridesharing FHV in NYC and the Manhattan Congestion Surcharge. Once quantified, the mode-shift impacts can be translated into the economic, environmental, societal impacts of the considered scenarios, aiming to quantitatively inform stakeholders and policymakers of the implications of shared mobility and congestion pricing on the entire city as well as specific populations and neighborhoods.

\section*{Data and Methods}

\subsection*{Data}
The data for the six major transportation modes in question (transit, walking, driving as well as taxi, FHV, shared FHV) is leveraged from three major sources: 1) C2SMART simulation test bed, 2) NYC Taxi and Limousine Commission (TLC) and 3) web-scraped data from public API interfaces. The C2SMART simulation test bed \cite{he2020pricing} includes approximately 27.3 million trips for travel modes - taxi, transit, walking and driving and across 16 income groups, following the travel agendas from the historic Regional Household Travel Survey (RHTS) with synthetic population. The data provides a representative estimation of the city-wide travel choices during pre-FHV era across people from different income groups across NYC. Whereas for estimation of taxi and FHV choices, we use the up-to-date open data from TLC. It is further used for estimating time and costs estimates for taxis and driving. The travel costs and times for other travel modes are retrieved from the publicly available API services (Google Maps and/or HERE Maps). Accounting for uncertainty is the key to our analysis, so we retrieved this information multiple times for each origin-destination (O-D) pair in order to capture the variations in the costs and times. More details regarding each data set are given in the Appendix A: Data \nameref{Appendix_data}.

\subsection*{Methodology}

Our objective is to prototype a simulation modeling framework suitable for understanding the mode-choice behaviour and assessment of city-scale impacts of transportation innovations and policies on urban transportation systems along with the associated environmental, economic and social implications. The assessment will be evaluated on two pilot use cases of introducing ride-sharing in NYC (offered through UberPOOL, Lyft Shared and other FHV companies) as well as Manhattan Congestion Surcharge. The impacts in question include: travel time and cost for passengers, traffic and congestion, gas consumption/vehicular emissions. Particular focus will be made on the equitability of the impact across populations. 

Traditional counterfactual impact assessment is challenged by 1) the fact that spatial counterfactual does not seem feasible (interventions are implemented city-wide and there is no comparable territory without deployment to be considered as control area), while 2) utility of the temporal counterfactual (comparing the same urban system before and after the deployment) is limited by multiple major trends and transformations happening within a complex urban system simultaneously with the deployment in question, 3) many target quantities of interest, such as overall urban traffic, gas consumption, emissions are hardly measurable with the available data and are again affected by multiple urban transformations happening simultaneously. 

As an alternative, this study proposes a methodology based on a data-driven integrated transportation simulation modelling framework, assessing the mode choice between walking, private and public transportation, taxi and FHV, including ride-share modes. For an estimated transportation demand, an agent-based choice model will be simulated, estimating unknown parameters of the individual utility of considered transportation modes as well as the agent characteristics (distribution parameters for individual preferences) through a multi-step Bayesian inference framework sequentially gaining information from the available partial observations of actual mobility choices. The Bayesian inference framework for the mode-choice model inference was earlier applied in our work on assessing the impact of bike-sharing \cite{Sobolevsky2018}, although the model used there was a more traditional multinomial logit discussed below.

While the simulated individual choices within the model enable direct assessment of the mode-shift consistent with individual preferences, which can be further translated into the impact of interest. Uncertainty of the data and parameter estimates will be incorporated into the simulations and resulting impact assessment.

\subsection*{Multinomial Logit}

We first consider the broadly used Multinomial Logit (MNL) model as the baseline approach for estimating the mode-choice for the regular commute. The model as well as its nested version (which we can use in case of related modes like taxi and FHV) offers an advantage of estimating the mode-choice probabilities using closed-form formulas representing the aggregate-level choices of a simulation model. However the parameters of the nested model lack the direct connection with the underlying simulation parameters and this way limit the utility of the model for individual-level mode-shift assessment. Nevertheless it can still serve as a baseline to assess efficiency of the proposed simulation model, so we include it in that capacity.

\begin{figure*}[!htbp]
    \centering
    \includegraphics[width=0.5\linewidth]{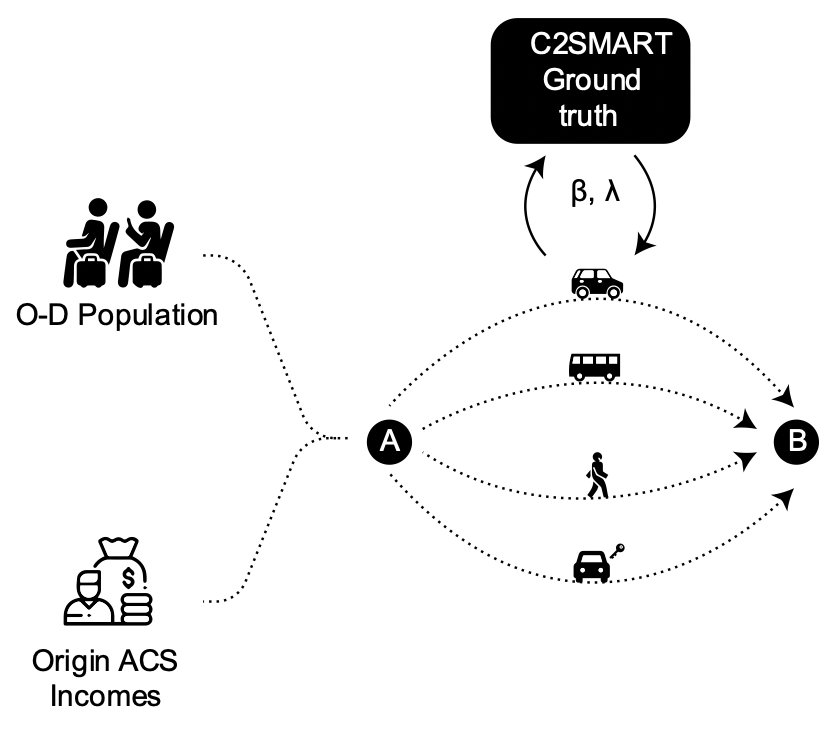}
    \begin{center}
        \textbf{Fig. 1 Multinomial Logit model framework}
    \end{center}
\end{figure*}

A MNL discrete choice model (Fig. 1) and its nested version with a nest for taxi and FHV and sub-nest for shared and non-shared FHV are trained based on the two available datasets: 1) number of trips between each O-D pair by wage group and 4 transport modes (taxi, transit, walking, driving) from C2SMART simulation test bed; and 2) number of trips between each O-D pair by 3 transport modes (taxi, FHV, shared FHV) from TLC. The models depend on a set of parameters - first of all $\lambda$, controlling the impact of the mode utility differences on the mode choice probability, $\beta$ adjusting the objective value of time (time multiplied by individual wage rate) to anticipated monetary cost incorporating possible irrationality of individual decisions while combining it with the direct monetary cost in order to assess the overall utility. The nested model would further include $\tau_{taxi+FHV}$, $\tau_{FHV}$ controlling the choices between nests and within each nest \cite{koppelman2006choice}.

Mathematically, the utility score $U_j$ for alternative \emph{j} depends on the time taken $T_j$ between the O-D pair in consideration, the monetary cost $P_j$ for choosing the alternative, the hourly income $W$ of the commuter, and a random component of error $\epsilon_j$, yielding a base utility function
\begin{equation}
    U_j = -(\beta WT_j + P_j)
\end{equation}
and the individual utility of $U_j+\epsilon_j$. If $\epsilon_j$ follows a Gumbel distribution it can be seen and that the probabilities for each of the four major transportation modes to be chosen as having the highest utility is defined as
\begin{equation}
    P_{mode} = \frac{e^{\lambda U_{mode}}}{e^{\lambda U_{taxi}} + e^{\lambda U_{transit}} + e^{\lambda U_{walk}} + e^{\lambda U_{drive}}}
\end{equation}

We further consider another version of the MNL with log-utilities (logMNL), corresponding to having multiplicative random factor applied to original utilities. Specifically adjust Equation (1) as
\begin{equation}
    U_j = -(\ln(\beta WT_j + P_j))
\end{equation}
considering log-utilities and assuming individual log-utility to be $U_j+\epsilon_j$ with a random term again following Gumbel distribution. This will correspond to choosing a mode with a minimal inverse utility $e^{-U_j}=(\beta WT_j + P_j)e^{-\epsilon_j}$ rather than a minimal negative utility $-U_j=\beta WT_j + P_j-\epsilon_j$ in the classical setup, i.e. having a multiplicative exp-Gumbel individual random factor instead of an additive Gumbel random term.

When considering FHV and shared FHV modes one needs to acknowledge the relation with the taxi mode and corresponding correlations between individual preferences. This is accounted by introducing a nest of taxi and modes along with a subnest of FHV and shared FHV modes to the model. For the nested model, the marginal probability of the outcome \emph{j} is calculated based on the deterministic part $V_j$ of the utility (i.e., $V_j = -\lambda(\beta WT_j + P_j)$), and the inclusive value $IV_k$ which signifies how inclusive each nest is based on its dissimilarity parameters (i.e., $IV_k=\ln \sum_{l \in N_k}e^{\frac{1}{\tau_k}V_l}$), yielding a chosen mode
\begin{equation}
    P_r(y=j) = \frac{e^{\frac{1}{\tau_k}V_j}}{e^{IV_k}}\cdot \frac{e^{\tau_kV_j}}{\sum_m \tau_m IV_k}
\end{equation}

The parameter $\tau_k$ cancels itself out for the nests containing a single transport mode. Eventually, the dissimilarity parameters $\tau_{taxi+FHV}, \tau_{FHV}$ for the taxi, non-shared FHV, shared FHV nests/sub-nests together with the utility parameters $\lambda, \beta$ determine the shift between each alternative, while $\tau_{taxi+FHV}, \tau_{FHV}$ largely control the balance within the taxi-FHV nest and FHV sub-nest.

The baseline model parameters were estimated through estimating $\lambda, \beta$ of the utility function based on C2SMART simulation test bed data by minimizing the Weighted Root Mean Squared Error (WRMSE)  between the number of trips from model prediction and real data for taxi, public transit, walking and driving. The tested models measure the goodness of fit between model prediction and ground truth data based on several metrics and search a wide range of parameters for the optimal fit in reasonable time. The final nested model splits commuters' regular mobility between origins and destinations across the city (from C2SMART simulation test bed or LEHD data) and predicts aggregated transportation mode choices for each origin in high consistence with C2SMART simulation test bed or ACS data. The model also provides wage distribution for each transport mode to be used while assessing preferred transport mode choice for the commuters from the given wage group. In the next evolution of the model it will enable further direct simulation of their future choices under changing conditions according to the scenarios of interest. 

\clearpage
\subsection*{Individual choice based simulation model}

This approach is based on agent-based simulations of individual choices. In fact so does the MNL model representing one particular scenario when the individual preferences are represented by an additive random term following Gumbel distribution. This enables a closed form representation of the resulting probabilities, however not relying on that allows further flexibility in choosing the modeling framework. Besides, direct control of the original simulation parameters will enable direct individual-level assessment of the mode-shift consistent with individual preferences.

\begin{figure*}[!htbp]
    \centering
    \includegraphics[width=0.88\linewidth]{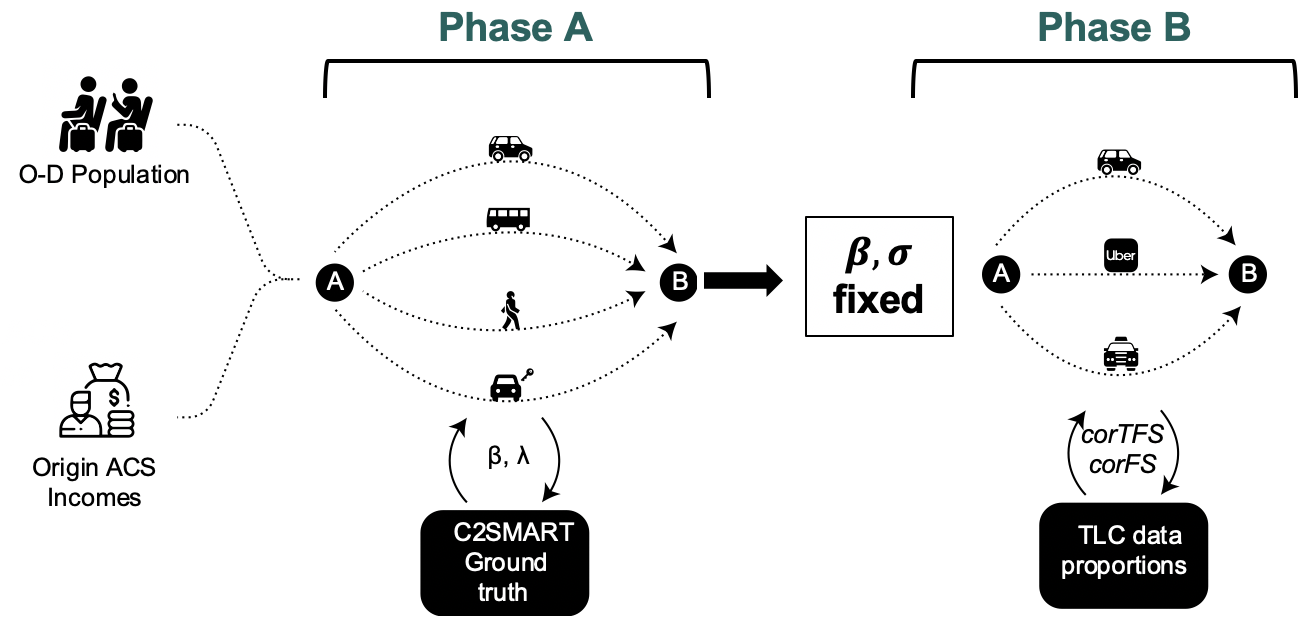}
    \begin{center}
        \textbf{Fig. 2 Individual choice based simulation model framework}
    \end{center}
\end{figure*}

In general this model simulates mode choices for each individual O-D pair and the specific passenger of the given income category. We will use the two-stage Bayesian inference framework based on the the data of individual simulated trips generated by C2SMART simulation test bed as well taxi and FHV data available from TLC in order to estimate the model (Fig. 2). Specifically, for each set of the model parameters we can simulate the choices for each O-D pair based on available transport modes. For each given pair of O-D, passenger wage and transportation mode, we assess utility based on travel time and cost estimates as well as a random factor, representing individual preferences towards each mode. Then log-utility is defined as $\ln U = \ln(\beta t w + c) + \epsilon$, where $\beta$ is the rationality adjustment for cost of time estimate as before, $t$ is the travel time estimate, $c$ is the travel fare/cost estimate and $w$ is the wage of the commuter, while $\epsilon \sim N(0,\sigma^2) $ is the random component representing individual preference to the given mode. We assume $\epsilon$ terms generally independent across transport modes except of taxi, FHV and shared FHV, which are of course related - if one has increased preference towards taxi, its likely that FHV will be also preferred and even more so between FHV and shared FHV which one can see as even more closely related, as while offering a slightly different type of service they are facilitated by the same provider/app. This way the model parameters to be estimated are: $\beta$, $\sigma$, $corTFS$ - correlation coefficient between random factors $\epsilon$ of taxi and FHV or shared FHV (SFHV) modes and $corFS$ - correlation coefficient between random factors $\epsilon$ of FHV and SFHV. 
 
\subsubsection*{Likelihood estimation - probabilistic approach}

Unlike the classical MNL model setting, the proposed model might not have a closed-form analytic solution for the choice probabilities. While estimating those through simulations might be computationally costly. In order to streamline the computational process we implement a neural network (NN) model fitting the mode-choice probabilities $P_m$ between the six transportation modes $m$ as the function of their log-utilities and the model parameters. Notice that $\sigma$ can be treated as the scaling factor for the log-utilities in order to simplify the model. In order to fit the model we simulate $P_m$ for various values of of $U_m/\sigma$ sampled from the random (normal) distribution and $corTFS$, $corFS$ (provided $corFS>corTFS$) sampled uniformly and use it to learn the neural network over 50,000 random trip samples. The model architecture consists of three hidden layers with 8,12,8 neurons respectively, with a rectified linear unit (``relu") activation for hidden and sigmoid for the output layer trained on ``binary cross-entropy" objective function. We further use this pre-trained NN model for estimating $P_m$ and further computing the likelihood of the observed C2SMART simulation test bed and TLC data given the model for each set of $\beta, \sigma$ and correlations parameters instead of using simulations or explicit analytic formulas.

For mode choice probabilities $P_m(o,d,w,\sigma,\beta)$ for each set of origin($o$), destination($d$) and wages($w$), the log-likelihood for four modes given the observed C2SMART simulation test bed ridership $R_m(o,d,w)$ is calculated as
\begin{equation}
    L(\sigma, \beta) = \sum_{o,d,w}\sum_m R_m(o,d,w)\ln P_m(o,d,w,\sigma,\beta)
\end{equation}
For each $\sigma, \beta$ the observed TLC ridership $R_m(o,d,w)$ for $m \in {taxi,FHV,SFHV}$ estimated using $P_m(o,d,w,\sigma,\beta,corTFS,corFS)$ calculate the log-likelihood of the data given the model as
\begin{equation}
    L_{FHV}(corTFS,corFS) = \sum_{o,d,w}\sum_m R_m(o,d,w)\ln \frac{P_m(o,d,w,\sigma,\beta)}{P_{TFHV}(o,d,w,\sigma,\beta)}
\end{equation}
where $P_{TFHV} = \sum_{m \in {taxi,FHV,SFHV}}P_m$. 

Based on the above framework, we obtained the best parameter sets of $\beta = 0.71, \sigma = 0.38$ and $corTFS = 0.31$, $corFS = 0.58$ based on likelihood values of the two stages. The parameter values $\beta, \sigma$ are sampled from log-normal prior distributions with $\ln\beta \sim N(\ln\mu_{beta}, \sigma_{beta}^2)$ and $\ln\sigma \sim N(\ln\mu_{sigma}, \sigma_{sigma}^2)$. The prior assumes having majority of the time underestimated up to $3$ times with $P(0.33<\beta<1)=68\%$ confidence, i.e. $P(-\ln 3<\ln\beta<0)=68\%$ which can be achieved when $\ln\mu_{beta}=-(\ln3)/2$ and $\sigma_{beta}=(\ln 3)/2$. Similarly, for $\sigma$, the prior distribution assumes having the individual correction factor $\epsilon$ within $[1/2,2]$ (correction up to twice) with $68\%$ confidence. This can be achieved if we take $\mu_{sigma}=\ln(\ln2)$ and $\sigma_{sigma} = \abs{\ln(\ln2)}$; if one simulates multiple $\ln\sigma \sim N(\ln(\ln2), (\ln(\ln2))^2)$ then for the resulting $\epsilon$ the probability of $P(0.5<\epsilon<2)$ is again going to be 68\%. The correlation parameters $corTFS$ and $corFS$ are sampled from uniform distribution [0,1] provided that $corFS>corTFS$. Then the sampling simply takes the evenly distributed percentiles of each distribution with equal weights.

Once the parameters are sampled and the model fit likelihoods are assessed, it allows simulating the mode-choices for a variety of sampled parameters with the results weighted by the joint likelihood $e^{L(\sigma,\beta)+L_{FHV}(corTFS,corFS)}$ (as the prior sampling ensures even probability intervals). For express-assessment one can simulate the results just for the max-likelihood parameters, however comprehensive parameter sampling provides assessment with respect to the model uncertainty. 

Based on the estimated parameter likelihoods, we simulate the final mode choices between origins and destinations for each individual commuter or group of commuters of a given wage group under two different scenarios of interest: A) intervention scenario having shared FHV unavailable or after imposing Manhattan Congestion Surcharge and B) the baseline scenario with all the transportation modes available with their original utilities. Individual correction factors $\epsilon$ are maintained the same between scenarios A and B. For each individual simulation and the set of model parameters the mode-shift can be directly assessed and aggregated into percentage mode-shift over the entire city or origin, destination and/or wage group of interest. Being assessed for multiple sampled parameters, it also provides probability distributions with respect to parameter likelihood weighting. The percentage mode-shift can be further translated into the impacts of interest with respect to the differences in travel time, cost and mileage driven between the transport modes.

\subsection*{Model comparisons}

We first evaluate the above simulation model against the classic MNL and logMNL (a version with multiplicative random factors for further consistency with the simulation framework above) according to their capability of fitting the reported choices of four major modes (walking, driving, public transit and taxi) during the pre-FHV era.

All of the discussed approaches estimate mode-choice probabilities $P_t$ for each origin-destination-wage pair based on the defined utility involving income of a commuter, travel time and costs. The MNL framework gives probabilities based on Equation (2). Whereas for individual choice simulation model, we estimate choice probabilities and resulting likelihoods for each parameter sets through a NN model. The simulations corresponding to each parameter set are weighted by the likelihoods having their logarithms estimated by (5) and (6). Table 1 reports the likelihood-weighted averages for the mode-choices provided by each model as well as the R-squared ($R^2$) values based on the net 4-mode prediction values for the models discussed.

\vspace{0.5em}
\begin{table*}[!ht]
    \begin{center}
    \textbf{Table 1 Comparison of aggregated number of modal trips of MNL vs Choice simulation model for 4 modes}
    \end{center}
    \centering
    \begin{tabular}{c|c|c|c|c|c}
    \hline
         & Taxi & Public Transit & Walking & Driving & $R^2$ score\\
    \hline
        Ground truth & 634,535 & 10,619,997 & 9,416,078 & 6,663,358	& -  \\
        MNL (multiplicative) & 409,560 & 9,920,182 & 8,038,709 & 8,957,035 & 0.870 \\
        MNL (additive) & 1,013,408 & 11,620,805 & 6,216,792 & 8,482,963 & 0.752 \\
        Choice simulation model & 692,391 & 8,781,046 & 8,533,870 & 8,001,092 & 0.899\\
    \hline
    \end{tabular}
    \label{tab:1}
\end{table*}

We observe that both multiplicative model specifications provide estimates much closer overall to the ground truth according to the $R^2$ score compared to the additive MNL specification, while individual choice simulation model performs slightly better compared to logMNL. But it also apparently gets a much closer prediction on the taxi ridership, which is particularly important for our use cases. Specifically, for the taxi ridership estimates (which is the most important for the considered use cases concerning taxi and FHV trips primarily), MNL underestimates the ground truth by over 1.5 times, while $\log$MNL overestimates by approximately 1.6 times. While the individual choice simulation model shows just a 9\% deviation. It also gives much closer estimates for walking and driving, while under-performing on the public transit. Furthermore, choice simulation model provides a more adequate estimate for the travel time rationality parameter (the max-likelihood parameter of $\beta=0.71$ corresponds to a quite realistic 29\% undervaluing time, while optimal $\beta$ for MNL and $\log$MNL is above $1$ corresponding to time overestimation, which contradicts common intuition of people generally valuing direct money benefits more than indirect benefits of the same estimated value. This further asserts that the main advantage of simulation model lies not in being vastly better than MNL evaluated on the whole data, but being more interpretable and providing better understanding of the underlying parameters and apparently a better fit as far as the taxi ridership is concerned.

Finally, as discussed the simulation model framework provides better intuition and flexibility when simulating individual trips and evaluating alternative choices for the mode-shift part of the analysis. Based on this initial evaluation we are going to stick to the simulation model going forward.

\subsection*{Uncertainty analysis}

Accounting for uncertainties is critically important for impact assessment in order to assess statistical significance of the reported city-wide quantities as well as their difference per wage group or areas across the city. We address uncertainties from two sources: 1) uncertainty in the data, and 2) uncertainty in the model. Uncertainty in the data is accounted for by incorporating the travel time and fares random distributions into the model and running the simulations multiple times. Model based uncertainty is analysed using the approach described above weighting results from different model simulations by the model fit likelihood. The variation in the trips from the data-based uncertainty simulations were observed to be pretty low to have any significant impact on the mode-shift and resulting impacts of interest. Whereas, this way uncertainty in the mode-choice assessment turns out to be much more significant (Appendix A: Uncertainty Analysis\nameref{Appendix_tables}). Hence going forward we primarily focus on this type of uncertainty in the mode-shift and related impact assessment. 

\section*{Impact Assessment}
In order to evaluate applicability of the proposed framework to assessing impacts of transportation interventions and policies, this study considers two use cases - introducing shared FHV after 2014 in NYC and imposing Manhattan Congestion Surcharge in early 2019.

\subsection*{Impact of ridesharing in NYC}

As shared FHV became an integral part of NYC transportation, understanding their actual impact is challenged by the lack of an appropriate control area where shared FHV were not available. Historic pre-2014 mobility cannot serve as an adequate baseline as a rapidly evolving transportation system likely got affected by multiple trends, not only the spread of shared FHV. E.g. increased adoption of an FHV service as such (not necessarily shared) could have had a larger impact.

However, the proposed mode-choice model allows simulating a hypothetical scenario with the same transportation demand if shared FHV were not available. As described before we first train the model on the historic mobility represented by C2SMART simulation test bed and then further estimate FHV-related parameters based on the actual taxi, FHV and shared FHV ridership reported by TLC. Important to mention that the model is used to simulate the relative distribution of the ridership per mode for each origin-destination and passenger wage group, while in order to estimate the actual scale of the impact we are going to rely on the actual amount of shared FHV reported by TLC (as those are the trips that would not have happened without ridesharing, while the alternative modes that would have been used are to be determined for those). This way dependence of the model on historic simulation test bed data is limited to estimating the likelihood of the parameters.

We analyzed the mode-shift (if shared FHV trips were to be facilitated by the second-choice mode in each individual scenario) simulated by the model with different parameters weighted by the model fit likelihood in order to determine the anticipated effect of shared FHV on the NYC transportation system. The mode-shift  (i.e. percentage of the observed shared FHV trips that would have been facilitated by public transportation, walking, taxi, FHV, and private vehicles) is reported on the Fig. 3. As one would expect majority of the shared FHV trips would have been facilitated by FHV and taxi as the closest alternative. Together with driving this adds up to nearly 70\%. However around 30\% of the trips have actually replaced transit and walking. So while majority of the shared FHV rides potentially (in case ridesharing actually occurred) cut the traffic by combining the trips that would otherwise involve individual driving, around 30\% of those trips replace non-driving mobility, this way increasing the traffic. The model-based uncertainties seem relatively small, highlighting robustness of the pattern. 

\begin{figure*}[!htbp]
    \centering
    \includegraphics[width=0.6\linewidth]{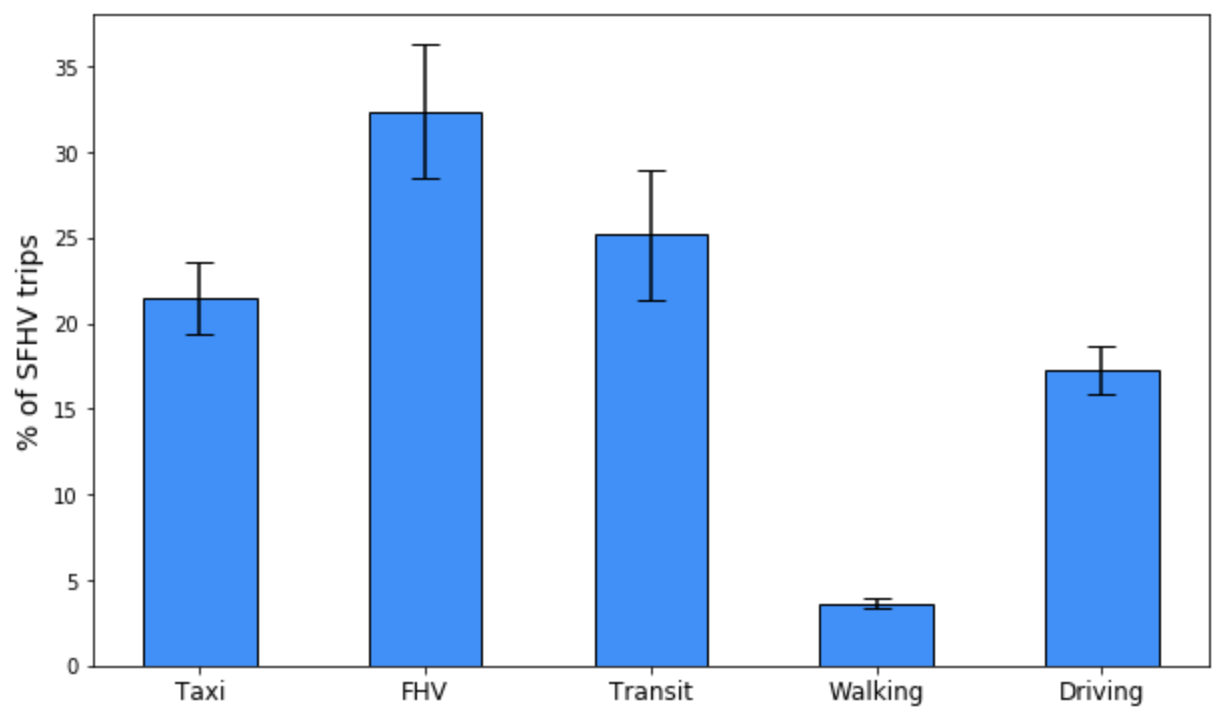}
    \begin{center}
        \textbf{Fig. 3 Percent of shared FHV trips accommodated by each alternative mode if shared FHV were not available}
    \end{center}
\end{figure*}

On the aggregate citywide scale we observe a net travel time decrease of $1.77\%$ ($95\%$ confidence interval [$1.71\%,1.83\%$]) and the net mileage increase of $1.14\%$ ($95\%$ confidence interval [$1.06\%,1.22\%$]) even if we assume that each shared FHV trip have actually combined two trips (unfortunately we do not have ground truth data on that, so this likely represents an optimistic scenario in terms of the traffic impact as some shared FHV might still serve individual passengers, while sharing more than two trips at once seems to be a rather rare case). Assessing on the scale of yearly citywide ridership from 2019, this corresponds to more than 495,000 hours saved for the NYC commuters at the price of 940,000 extra miles driven citywide over the year. So on average every hour saved comes at a price of a traffic increase by 1.9 miles. The net decrease in travel times comes mainly from the reduction of about 14M transit trips. The extra miles driven translate to close to 47,000 extra gallons of fuel emitting around 375 tons of carbon-dioxide emissions. In terms of economic impacts, the mode shift accounts for the citywide time-cost reduction of \$4.72M.

While shared FHV cause an overall travel time decrease and traffic increase across the city, those impacts are greatly uneven across the city. On the level of individual taxi zones, the largest travel time decrease of up to $8\%$ occurred in inner areas of Brooklyn, Queens and Staten Island which seem to benefit the most (Fig. 4) as the new relatively affordable commute option has likely bridged the local gaps in transportation accessibility. While some areas such as the airports actually saw an opposite effect of up to $8\%$ increase in travel time, which can be related to using the shared FHV as a replacement for more expensive taxi and FHV service heavily used in such locations (having generally lengthy and expensive commutes for which people may compromise travel time for a significant cost savings).

\begin{figure*}
\begin{minipage}{.5\textwidth}
  \centering
  \includegraphics[width=1\linewidth]{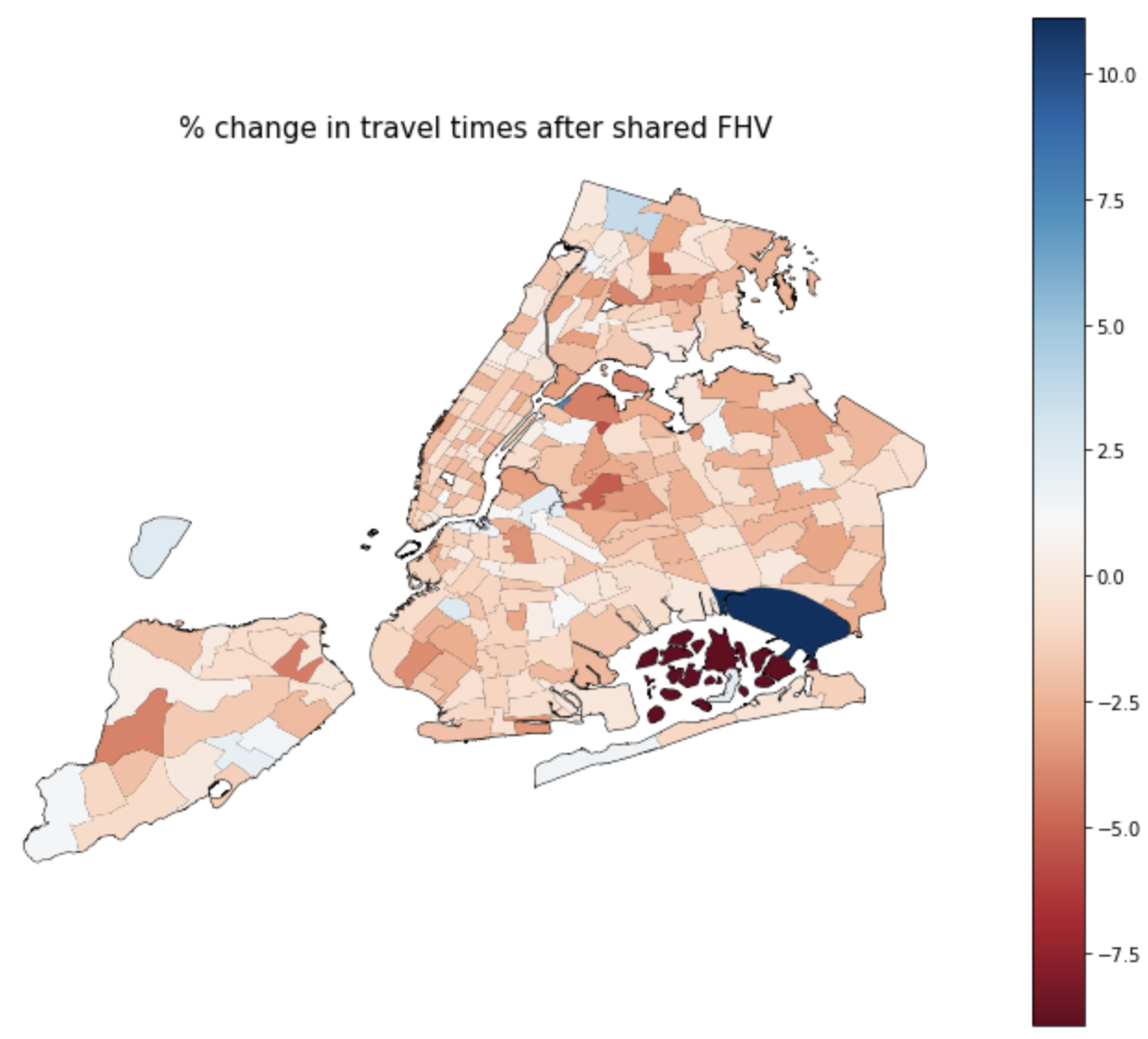}
  \label{fig:test1}
\end{minipage}%
\begin{minipage}{.5\textwidth}
  \centering
  \includegraphics[width=1\linewidth]{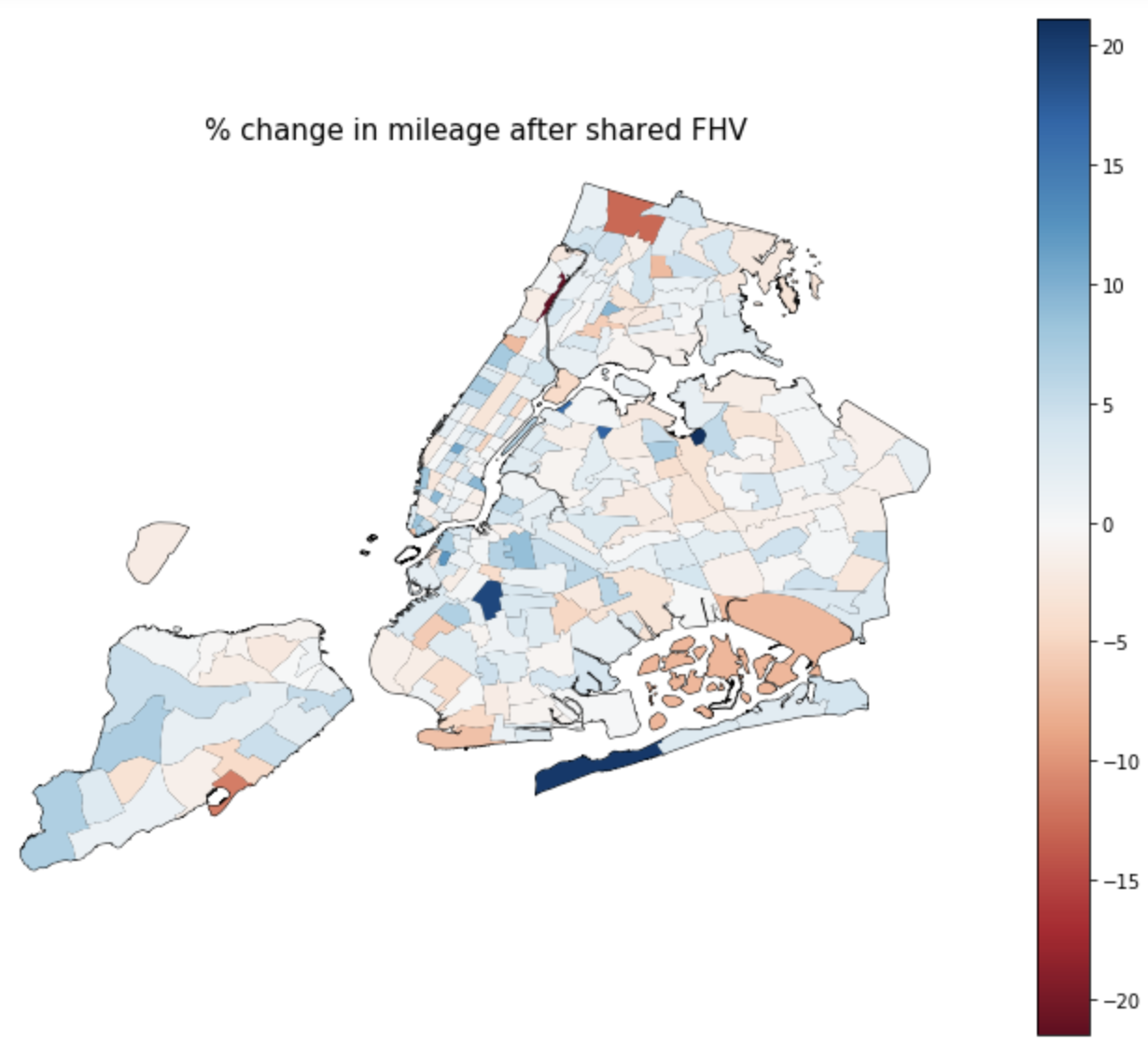}
  \label{fig:test2}
\end{minipage}
\centering
      \begin{center}
        \textbf{Fig. 4 Percent change in travel times and mileage across taxi zones}
    \end{center}
\end{figure*}

Providing individual simulations with respect to the commuter wealth, the model allows to analyze the equitability of the impacts across urban populations. We observe the most significant changes for the low income groups in \% difference in mileage (Fig. 5), while the highest changes in travel times are observed for higher income groups ($>$\$100k annual income). For the high income groups, the majority of shared FHV trips come from transit and driving. So there is an increase in mileage from transit to shared FHV trips and at the same time decrease from switch from driving to shared FHV. In case of low income groups ($<$\$60k annual income), the mileage increase comes from shared FHV trips are being accommodated from walking and transit modes. In short it looks like the shared FHV service is the most efficient for the wealthier in terms of the trade-off between improved travel time and the traffic footprint, while when used by low-income passengers it causes much heavier traffic footprint with smaller travel time improvement. Additionally, we observe that the mode-shift differences across income groups are significant with respect to the model-based uncertainties.

\begin{figure*}[!htbp]
    \centering
    \includegraphics[width=0.85\linewidth]{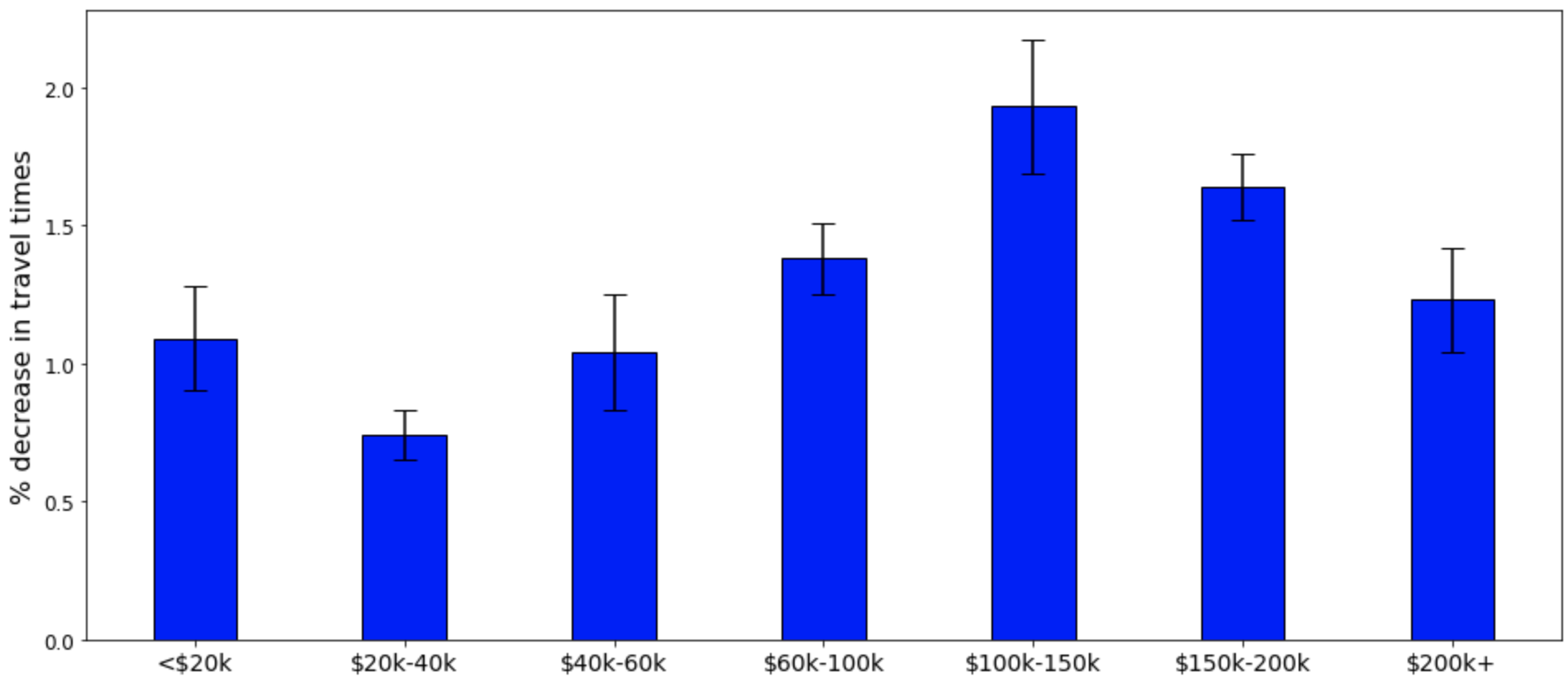}
    \centering
    \includegraphics[width=0.85\linewidth]{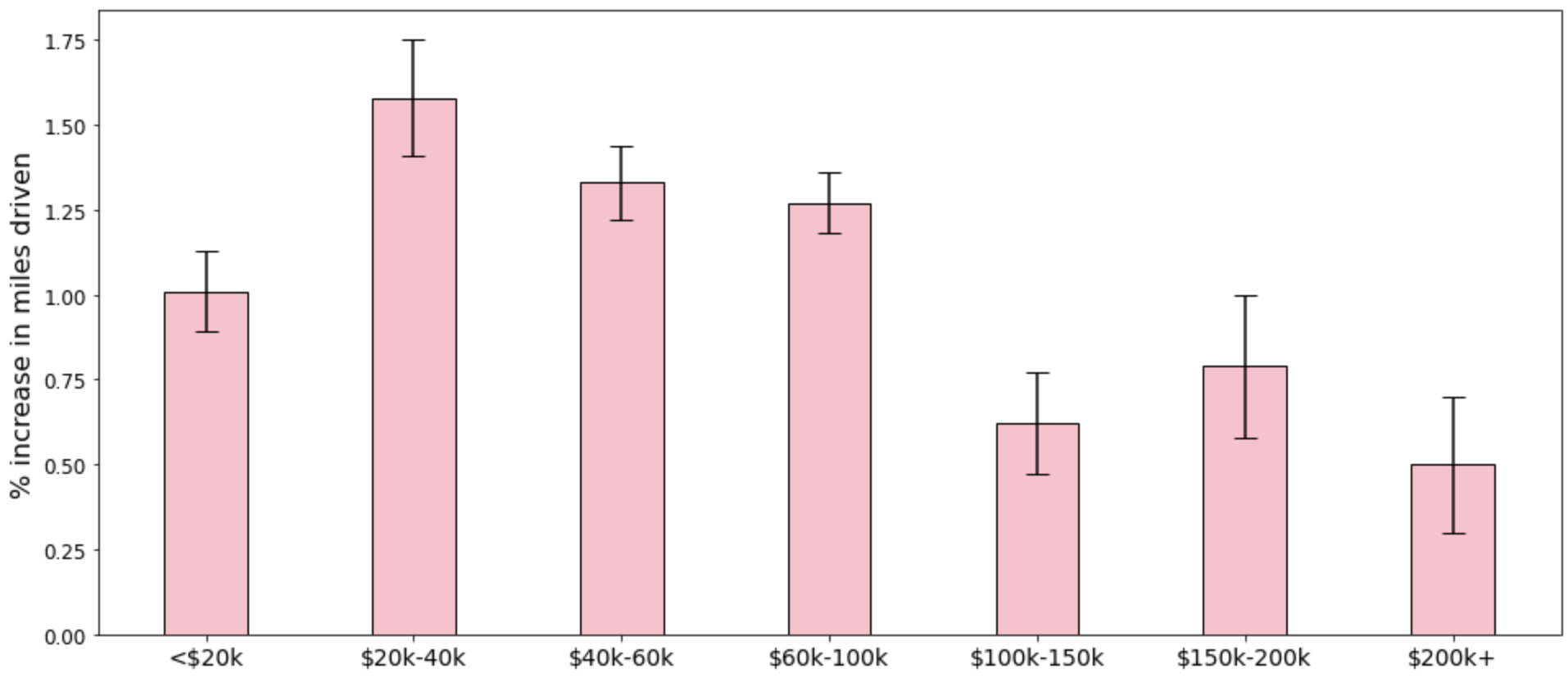}
    \begin{center}
        \textbf{Fig. 5 Percent changes in travel timess and mileage across income groups}
    \end{center}
\end{figure*} 

With the intervention scenario of introduction of shared FHV, we can expect an increase in citywide net travel mileage and decrease in travel times. But as the C2SMART simulation test bed data might not be a perfect representation of true mobility within the city, it is important to test our model across different mobility data sets to see how much the results might differ and if the model gives a reasonable estimate across different representations of mobility. Thus we decided to also test it on one other data for NYC– the Longitudinal Employer-Household Dynamics (LEHD) mobility. The LEHD data has mobility information from across 47,000 O-D pairs compared to $\sim$21,000 pairs from C2SMART simulation test bed. On running the simulation model with the best likelihood parameters on the LEHD pairs, we observed a net citywide travel time decrease of $1.91\%$ and a net mileage increase of $1.29\%$ upon introducing the scenario where shared FHV was available as a transport mode. So the resulting impacts from the simulation model mildly depend on the data source of mobility demand (and compared to other quantifiable sources of the assessment uncertainty, the data source remains the most significant one), but generally remain consistent and close to the range of percent changes originally obtained from the C2SMART simulation test bed data. 

\subsection*{Manhattan congestion pricing impact}

Another use case is the impact of a new pricing policy - Manhattan Congestion Surcharge, adding a fixed cost to taxis (\$2.50), FHV (\$2.75) and shared FHV (\$0.75/passenger) for all trips originating in Manhattan. For shared FHV, we took an average of 2 passengers per ride at a time, so the total cost added was \$1.50.

On seeing the number of reduced trips across modes, we observe an almost equal drop across taxis and FHVs, although the highest reduction is seen for shared FHV. Almost $60\%$ of the reduced trips are accommodated by transit mode, which translate into \$16M projected increase in revenue for the MTA. Driving and walking accommodate $28\%$ and $12\%$ of the reduced trips respectively (Fig. 6). Assessing on a scale of total 2019 taxi and FHV ridership, the decrease in number of trips for taxis and FHVs account for around 681,000 less miles driven which comes at a net increased travel time of 329,000 hours. The decrease in driving mileage causes revenue loss of \$19M for the taxi and \$11M for FHV (shared and non-shared) which comes due to the drop in taxi and FHV trip numbers for trips originating in Manhattan, although the net revenue increase for taxi and FHV services is \$119M from the increased prices per trip. This further translate into the citywide economic impact of \$2.7M time-cost value increase after the Manhattan Congestion Surcharge is added.

\begin{figure*}[!htbp]
    \centering
    \includegraphics[width=0.6\linewidth]{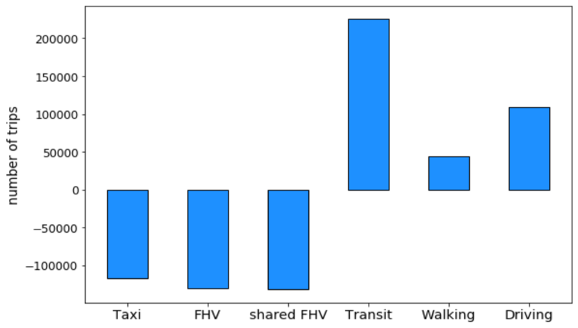}
    \begin{center}
        \textbf{Fig. 6 Change in number of trips across transport mode after Manhattan Congestion Surcharge (on C2SMART simulation test bed scale)}
    \end{center}
\end{figure*}

Seeing from an equitability perspective, we observe the most dramatic changes for the high-income groups in percent difference in travel times and mileage (Fig. 7), meaning that commute choices of the richest are affected the most. Compared to the low-income population, we see an increase of about $1\%$ in travel times for high-income groups. The same is observed for total mileage driven, where the decrease is about $0.8\%$ lower for low income populations than high income groups. This makes sense as taxi and FHV ridership is seen across high income population. The highest mileage cut comes from top mode switch from FHVs and taxi to transit. This change is seen the most for the \$100k-\$150k income group whereas for $>$\$150k income groups, the mileage cut decreases as top mode choice is private car instead of transit after the congestion surcharge. The mileage cut is significantly less for lower income groups as the number of trips of taxis and FHVs are low to begin with. With congestion surcharge, the top mode choice becomes transit/walking but the net number of trip changes are low compared to the higher income groups. The highest change among low income groups is observed for the \$60k-\$100k group where top mode choice switches to transit from non-shared/shared FHV after the congestion charge is introduced. The same trend is seen for the travel times where the rich observe highest time increase owing to their switch from taxis/FHVs to transit mode. Spatially, the biggest impact is seen in the high-income neighborhoods of Manhattan, specifically Lower East Side, Upper East Side and Upper West Side parts of the borough. As compared to Upper Manhattan neighborhoods like East Harlem and Washington Heights, the impacts in both travel times and mileage are relatively higher in Midtown and Lower Manhattan areas.

\begin{figure*}[!htbp]
    \centering
    \includegraphics[width=0.85\linewidth]{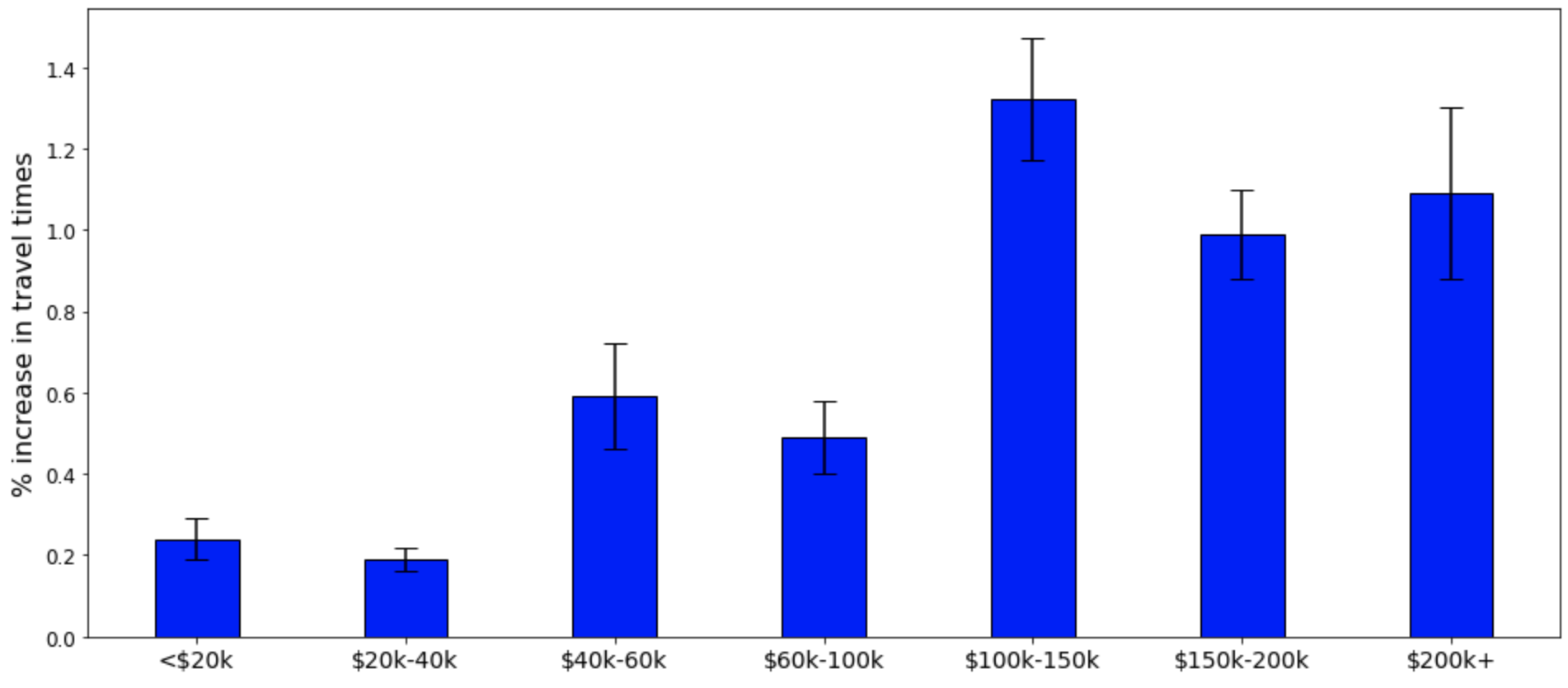}
    \centering
    \includegraphics[width=0.85\linewidth]{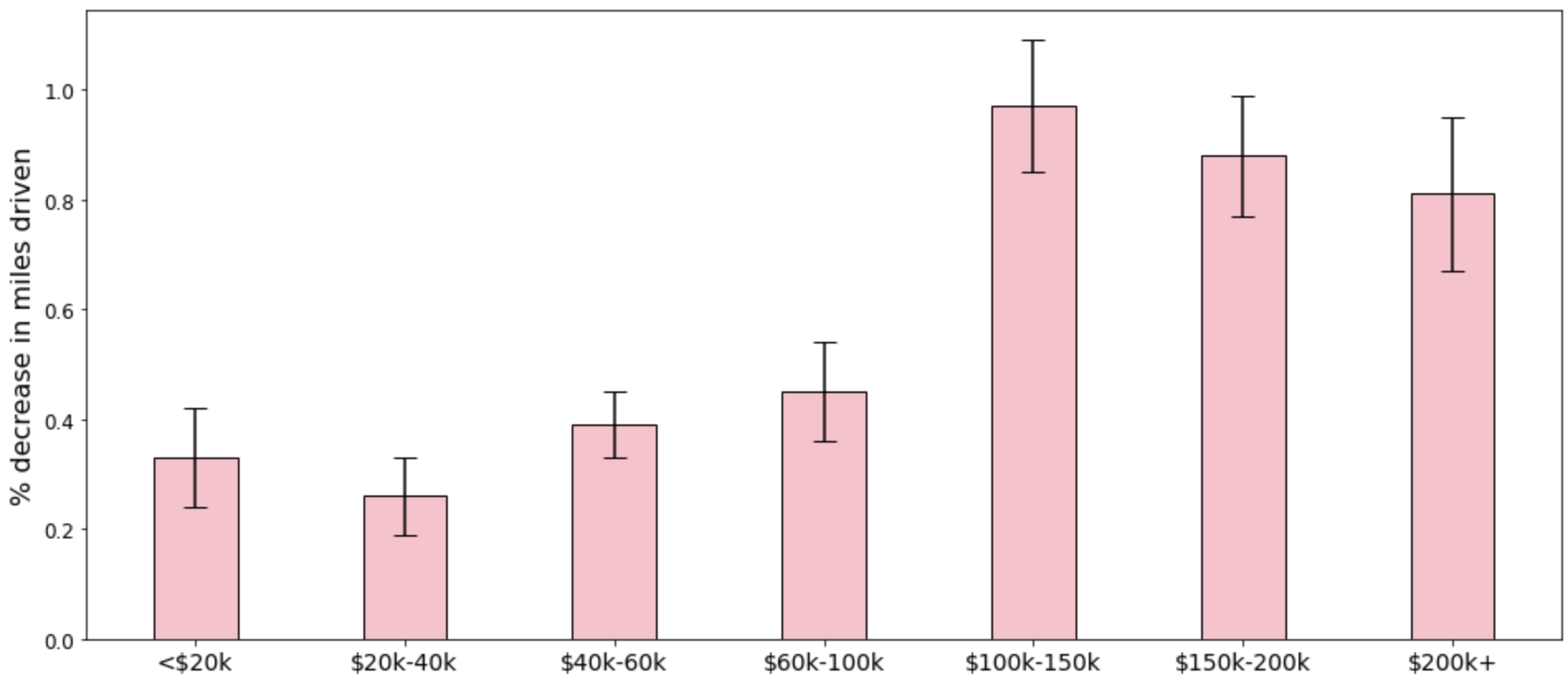}
    \begin{center}
        \textbf{Fig. 7 Percent changes in travel times and mileage across income groups after Manhattan Congestion Surcharge}
    \end{center}
\end{figure*}

According to the model simulation, on a city-wide scale we observe an increase of $1.09\%$ in travel times and a $0.87\%$ decrease in mileage, which can be attributed to lower usage of taxis and FHV and the mode-shift to alternative non-driving modes. The relatively low changes in total travel times and mileage for the whole city can be explained with the low total proportion of taxi trips present in the data. Together, taxis and FHVs make up to around $\sim7\%$ of the net mobility in the C2SMART simulations. Thus any monetary changes in fares in taxi and FHVs for one borough (Manhattan) translate into a low change in net times and mileages. So in general the policy seems to be efficient in causing a statistically significant decrease in the overall traffic, while the vulnerable populations seem to be the least affected overall.

\clearpage
\section*{Conclusions}

This study constructed the simulation modeling and probabilistic inference framework suitable for assessment of city-scale impacts of transportation innovations and policies on the transportation system along with the associated environmental and economic implications with respect to uncertainty of such impacts. The framework's applicability is illustrated on two use cases: introduction of shared FHV in NYC and Manhattan Congestion Surcharge. Also the framework is capable of learning from diverse and possibly inconsistent datasets (such as historic transportation surveys and actual taxi and FHV ridership) providing partial information on urban mobility,  stepwise gaining information from either source.

Broadly, our results indicate that shared mobility helped decreasing travel times between 1-2$\%$ for all categories of passengers. However, it does so by increasing the traffic up to 0.5-1.5$\%$ – decreases from trip sharing seem to be offset by growing number of riders due to increased affordability of the service. It works more efficiently for high-income categories of passengers providing higher travel time decrease with lower mileage increase. On the other hand, the Manhattan Congestion Surcharge noticeably decreases the FHV traffic of up to 1$\%$, however it does so at the price of increasing travel time and in particular for high income travelers, who are perhaps the most frequent users of taxis and FHVs, to which the surcharge is targeted. The uncertainty analysis confirms statistical significance of the impacts as well as their heterogeneity across populations. The impacts above are further translated into the total traffic, gas consumption, emissions, monetary savings and public transit earnings implications.

While we hope that this study can be a proof of concept for other cities considering shared mobility, congestion pricing or other similar interventions, it should be noted that NYC’s transportation system is unique in many ways, and makes a switch to public transportation more practical than in many other cities. In addition, while the impact assessments in the paper provide proof-of-concept use cases for the proposed framework, further work may be needed to develop a comprehensive and accurate picture of the mode choices and mode shift. The outdated survery-based ground truth refined by C2SMART simulation test bed by itself might not be fully representative to the actual urban mobility. The current landscape of urban mobility might differ significantly from the RHTS and similar available transportation surveys conducted in the pre-FHV era. And although the historic data is only used as part of the parameter estimation for the model, while the scale of the impacts is based on the up-to-date TLC data, the mode-choice proportions and reliability of the impact assessment might still get affected. 

Another limitation of the study is the simplicity of the utility function as presently considered. Accounting only for travel time and cost it may not reflect all the critical factors of how a person makes a transportation choice. The present utility function would work very well in an ideal world where everyone worked out the economics of their commute daily, but, the reality is that transportation choices are influenced by habits, comfort preferences, and other human factors as well as environmental conditions. We focused our study on commuters because it allowed us to infer demographic and transportation demand information, but the morning commute takes up a small part of NYC’s complex transportation system. The collection of more comprehensive ground truth data and accounting for more aspects of individual choices could further improve the reliability of the impact assessment. Finally, the validity of the mode-choice and impact assessments is conditional on the validity of the model, although an uncertainty assessment related to inaccuracies in the data as well as the model fit allows us to assess the degree of confidence in such an assessment, the specific model behind mode choices need to be assumed. 

With that, the main contribution of the paper is a proof-of-concept demonstration that a robust data-driven probabilistic modeling framework incorporating incomplete and inconsistent available mobility data, is capable of assessing the holistic picture of the urban commute and impact of transportation interventions with a reasonable degree of certainty.

\subsection*{Acknowledgements}

We thank Kaan Ozbay from NYU C2SMART, Satish Ukkusuri from Purdue University, Peter Glus, Yuan Shi from Arcadis as well as Patrick Smith from NYCDOT for helpful discussions. We also thank Brian Yueshuai He and Joseph Y. J. Chow from NYU C2SMART for sharing the C2SMART simulation test bed data. We further thank Chinmay Singhal and Mingyi He for their help with data processing and visualization. Chaogui Kang was supported by the National Key Research and Development Program of China (no. 2017YFB0503604) and the National Natural Science Foundation of China (no. 41601484 and 41830645). Stanislav Sobolevsky and Devashish Khulbe were supported by the USDOT (award no. 69A3551747124 through C2SMART USDOT Tier 1 Center).

\subsection*{Author contributions}
C.K. and S.S. designed research; D.K. and C.K. performed research; C.K., D.K., and S.S. analyzed data; C.K., D.K and S.S. wrote the paper.


\subsection*{Competing interests}
The authors declare no competing financial interests.

\clearpage


\clearpage
{\textbf{Appendix A}}\\

\section*{Data} 
\label{Appendix_data}

Data from following sources - C2SMART simulation test bed, the Regional Household Travel Survey (RHTS) and the NYC Taxi and Limousine Commission trip records (TLC) - was used to determine the transportation demand between O-D pairs and the wage distribution of commuters. Initial exploration of the RHTS/TLC data supported our choice to focus on the commute hours (i.e., 7am - 10am and 5pm - 8pm). 

\subsection*{C2SMART simulation test bed}

This data provides travel demand for 4 modes - taxi, transit, walking and driving. The 27.3 million trips are aggregated on Traffic Analysis Zones (TAZ) which are further aggregated into taxi zone levels for our models. In total, the data covers trips from 20,834 unique O-D pairs, covering 250 of 263 taxi zones in New York City (Appendix Fig. S1). The data also contains trips from modes like bike, carpool, shared bikes etc., which we do not include in our analyses. Nearly half of the trips constitute transit, followed by driving, walking and taxi. The data was simulated by following the travel agendas from the historic Regional Household Travel Survey (RHTS) with synthetic population \cite{he2020pricing}.

\textbf{Regional Household Travel Survey.} The data was used to reflect reported choices of transportation modes by commuters serving as partial ground truth for fitting the model. Census tract level estimate data was pulled in order to generate probabilities of a commuter within each taxi zone of having wages within each Census income bracket. Commute information from the RHTS was reported by respondents which form of transportation they used “most days” for commuting to work, as to estimate the percentage of people in each taxi zone that regularly choose each form of transportation for trips to work. Collectively, the RHTS data allows us to estimate the probability of any given resident of each origin zone choosing each distinct mode of transportation, given the commuter’s income.

\newpage
\subsection*{LEHD/ACS data.} 

Another data source is from the Longitudinal Employer-Household Dynamics (LEHD) program. It has commuter information for 11 wage groups on a taxi zone level. Also present is the population choices of transportation for taxi, walking, transit, driving, biking, carpool etc. from the American Community Survey (ACS) data. Information regarding FHV is missing, so only the four transport modes of our interest are present. Finally, since only the origin-based commuter information is present, we cannot explicitly have any true choices for an origin-destination pair.

\subsection*{TLC open data}

\textbf{Taxi trips.} New York City’s Taxi and Limousine Commission provides free access to their database of taxi trip data with ride-level granularity. These data give a sense of the high volume traffic areas in the city, as well as the distribution of trips by time of day. They are crucial to estimate current modal distribution and, when used in conjunction with demographic RHTS (or C2SMART simulation test bed) data, to predict mode shift under various scenarios and within varying demographics. We found that the actual TLC data was most correlated with the LEHD Origin-Destination Employment Statistics (LODES) demand for the commute hours, and this correlation allows us to assume that the extracted trips are largely originating from the rider’s home taxi zone, enabling us to infer the commute trips by taxi from TLC data between taxi zone pairs (O-D pairs). 

\textbf{For-Hire-Vehicles trips.} The data was also provided by TLC consists of individual trip data for different FHV services (ex. Uber, Lyft, etc.). For analysis, we have separated the FHV trips to FHV and shared FHV, and aggregated both data at the level of pick-up and drop-off zone, date and hours between commute hours. The aggregated data contains attributes of date, pick-up location id, drop-off location id, average trip duration (sec), trip counts, and surcharge flag (FHV and shared FHV). A typical month of data includes 15 to 20 million rides, with around $20\%$ of shared FHV, and $80\%$ non-shared FHV. We found the zones with high trip amounts are mostly concentrated at lower/middle Manhattan and downtown Brooklyn areas for both pick-up and drop-off locations. Such finding suggests that a large portion of our model simulations will be reflecting the trips in these areas, thus we need to be more carefully consider the regional demographic information of these areas as well as their functionalities (e.g. shopping, parks, etc), to avoid any false assumptions when profiling people choices. 

\subsection*{Travel times and costs}

The key elements of our model required us to estimate the time and cost associated with trips between each taxi zone pair  for each of the six transportation modes (taxi, FHV, shared FHV, public transportation, walking, private vehicle). Additionally, in order to evaluate the utility of each transportation mode, we obtained the rider’s wages from RHTS. We have aggregated all data sources to the TLC taxi zone level and the final version of the data which is used in the model includes pickup and drop-off locations, commute duration, price, and the wage distribution for that origin-destination pair. We considered mean fare amount, trip duration, and their standard deviations to inform data uncertainty.

\textbf{API services.} HERE and Google technologies are the company that provides mapping and location data. In order to assess travel time, cost and overall utility of each transportation mode considered in the model given the O-D pair, we use HERE/Google REST APIs to gather information such as maps, routing, geocoding, places, positioning, traffic, transit, and weather information. The public transit data can be acquired via specific Public Transit API. Using HTTP GET methods, route information such as the trip time duration, the number of transfer, and the mode for each transfer will be get given departure and arrive location, departure time, and specific mode. To try and limit the impact of any special circumstances that would impact these estimates, we retrieved the data on several occasions and took an average. For those pairs with no route information available for either of the modes, we consider their corresponding time and distance values to be infinity. This situation can happen where there does not exist a way of commuting from one zone to another (e.g. islands).

\newpage
\section*{Uncertainty analysis} 
\label{Appendix_uncertainty}

Uncertainty in the data is accounted for by incorporating the travel time and fares random distributions into the model and running the simulations multiple times. Then we calculate the mean and variance of trips for each of the four major modes from the simulations. 
The results of data-based uncertainty are in Appendix Table S1.

Model based uncertainty is analysed using the approach mentioned in the probabilistic approach of getting simulation likelihoods. For the log-normal prior distributions of $\beta, \sigma$ along with uniform distributions of correlation parameters (as described previously, $\ln\mu_{beta}=-(\ln3)/2$, $\sigma_{beta}=(\ln 3)/2$, $\ln\sigma \sim N(\ln(\ln2), (\ln(\ln2))^2)$, $corTFS$, $corFS$ $\sim$ [0,1]) , we simulate utilities $U_t(o,d,w)$ and weight by the likelihood values estimated by the formulas described in equations (4) and (5). We then estimate the weighted average and standard deviations of the simulations. We sample (10$\times$10) values of $\beta, \sigma$ and (10$\times$10) $corTFS$, $corFS$ pairs ($corFS>corTFS$)  drawn from the above distributions weighted equally. The 4-mode uncertainty results using this approach is given in Appendix Table S2.

\section*{Supplementary Tables} 
\label{Appendix_tables}

\begin{table*}[!ht]
    \begin{center}
    \textbf{Table S1 Data-based uncertainty results for four modes }
    \end{center}
    \centering
    \begin{tabular}{c|c|c|c|c}
    \hline
         & Taxi & Public Transit & Walking & Driving\\
    \hline
        Ground truth & 634,535 & 10,619,997 & 9,416,078 & 6,663,358	 \\
        Simulation averages & 692,391 & 8,781,046 & 8,533,870 & 8,001,092\\
        Simulation std & 2,241 & 4,449 & 5,480 & 5,808 \\
    \hline
    \end{tabular}
    \label{stab:1}
\end{table*}

\begin{table*}[!ht]
    \begin{center}
    \textbf{Table S2 Model-based 4-mode uncertainty results with probabilistic approach}
    \end{center}
    \centering
    \begin{tabular}{c|c|c|c|c}
    \hline
         & Taxi & Public Transit & Walking & Driving\\
    \hline
        Ground truth & 634,535 & 10,619,997 & 9,416,078 & 6,663,358	 \\
        Simulation weighted average & 692,391 & 8,781,046 & 8,533,870 & 8,001,092 \\
        Simulation weighted std & 105,438 & 509,274 & 496,029 & 459,127 \\
    \hline
    \end{tabular}
    \label{stab:2}
\end{table*}

\clearpage
\section*{Supplementary Figures} 
\label{Appendix_figures}

\begin{figure*}[!htbp]
\begin{minipage}{.5\textwidth}
  \centering
  \includegraphics[width=1\linewidth]{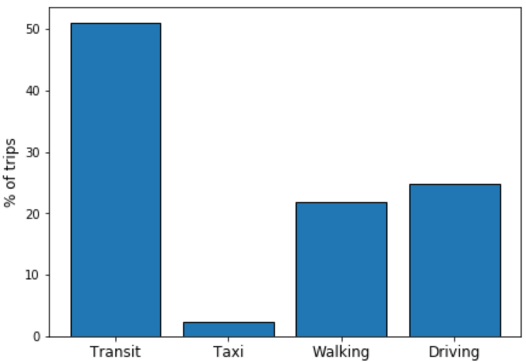}
  \label{fig:test1}
\end{minipage}%
\begin{minipage}{.5\textwidth}
  \centering
  \includegraphics[width=1\linewidth]{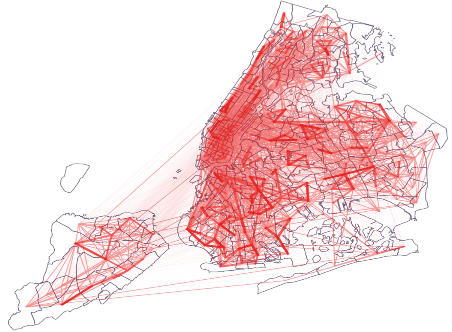}
  \label{fig:test2}
\end{minipage}
\centering
      \begin{center}
        \textbf{Fig. S1 Mode-wise distribution and taxi zone level distribution of C2SMART simulation test bed data}
    \end{center}
\end{figure*}

\begin{figure*}[!htbp]
    \centering
    \includegraphics[width=0.8\linewidth]{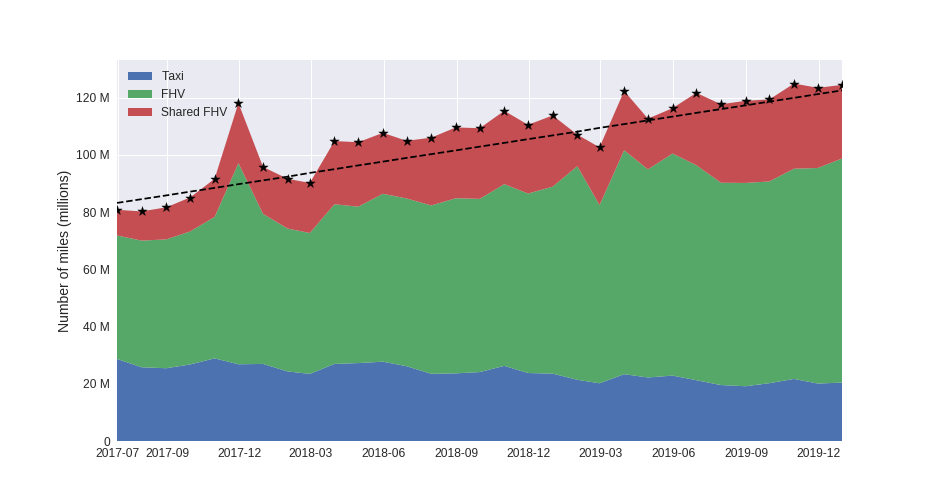}
    \begin{center}
        \textbf{Fig. S2 Net mileage and shares for taxi and FHVs in New York City (2017-2019)}
    \end{center}
\end{figure*}

\end{document}